# Spin-glass-like Dynamics of Social Networks

**Fariel Shafee**
*Physics Department*
*Princeton University*
*Princeton, NJ 08540*
*USA*

*fshafee@princeton.edu*

*Subject area*: econophysics, social and complex networks

**Abstract**

In this work we study spin-glass (SG) like behavior in the dynamics of multiple agents in a social or economic context using interactions which are similar to the physical case. The different preferences shown by individual agents are represented by orientations of spin-like variables. Because of limited resources, each agent tries to maximize her total utility function, giving a prescription for the dynamics of the system similar to the evolution resulting from the optimization of the interaction of a SG. The coupling between agents for different attributes may be positive or negative, as in a physical SG system, forming "frustrations" from the ensuing conflicts, with the system trying to find an overall equilibrium, but in vain, so that we observe oscillations. The couplings are provided by matrices corresponding to each attribute and each agent, which are allowed to have some fixed bias, indicating the unchangeable component of the make up of the agents from genetic factors or lasting environmental influences, and also contain a random part from environmental noise, i.e. the cumulative stochastic effect of lumped factors not explicitly accounted for in the model. In a simulation of a small world with a small number of agents and attributes we observe, for particular choices of the coupling matrices, interesting variations of behavior patterns, including oscillations with different long and short term behavior of punctuated equilibria. We also show that if the "spin" variables are extended to become co-ordinate-like variables, the interaction appears like the distance function in differential geometry, with the coupling matrices forming an indefinite metric in the state space of the system. We comment on the relevance and consequences of equations of motion that may be derived from such an analogy.

# I. Introduction

In economic and social contexts the individual agents show different preferences, which may be in part due to specific genetic make up and in part due to socio-cultural and educational background, i.e. lasting effects of the environment. On the other hand in economics we usually assume that agents behave rationally. Since preferences obviously differ, what appears to be rational to one agent may not be so to another. As each agent tries to maximize her satisfaction or utility function, she uses her perception of the other agents' preferences, which may not be what those agents themselves actually perceive. There may be deliberate bluffing also in a game theoretic [1] sense if agents consider that misleading other agents may be advantageous to them since resources are limited.

In a more general context, we have tried to explain the non-quantifiable human needs [2] and the diversity in utility functions by taking into account Gödel's incompleteness theorem [3]. Strategies to maximize satisfaction might be rational, but we have argued that the needs themselves might not be a "common rational" set of constants. As any logic system is based on a set of axioms that are by themselves not provable within the logic system itself, and the axioms or sets of information are acquired by each agent individually by interacting with its environment and other agents, this individuality will be reflected in the choices of each agent. Some needs, however, might be very common and many of these are quantifiable easily in terms of material or opportunity cost, whereas some other needs are almost impossible to express in terms of maneuverable quantities.

The utility function or the satisfaction derived from specific items as measured by the opportunity cost is based on the question of relative security of the 'self' obtained from the transaction, and as each agent tries to optimize her total satisfaction independently, with respect to some items it may be advantageous to align with other agents with similar orientations, and with respect to some others the choice for individual satisfaction may lead to opposite orientations, as in a minority game. Because of conflicting couplings of some attributes of the agents, there may be "frustration" as in a spin-glass model in physics [4, 5], with the system trying to find an overall equilibrium, but in vain, so that we observe oscillations of the orientations of the agents according to their individuality as well as their interaction with other agents, and also the linking weights between the agent pairs.

In the next section we first summarize the idea of a physical spin-glass, in section 3 we present our multi-agent social model analogous to a spin-glass system, in section 4 we formulate a procedure for the simulation of a small system and in section 5 we show the results of the simulation, in section 5 we outline a novel differential geometric picture of the interaction and discuss some of its consequences, and in section 6 we present our conclusions.

## II. Spin-Glass Models

The simplest soluble model in magnetism is the Ising model, which consists of a lattice of spins with only one component which can take the values "1" or "-1", i.e. which can only flip. The Hamiltonian is described by

$$H = -\sum_{i,k} J_{ik}\, s_i\, s_k - \sum_i h\, s_i \tag{1}$$

where the index *i* or *k* of the spin *s* is the label of the location site of the spin in the lattice, the first sum is over all nearest neighbor pairs, *J* being the nearest neighbor coupling strength of the spins, which is a constant and *h* is a constant external magnetic field. The Ising model is the simplest physical system that shows a nontrivial phase transition at nonzero temperature. The two-dimensional Ising model has been exactly solved by Onsager [6], but higher dimensions can only be simulated or approximated.

The spin-glass (SG) model describes a system with random coupling $J_{ij}$ between the spins $s_i$ and $s_j$, which can even change sign. So some links can behave as ferromagnetic and some as antiferromagnetic, and the inconsistency can produce "frustrations". Even a slight change in temperature can produce a totally different state with no overlap with the previous one. In a SG the order parameter is not the mean spin, which is zero, but the average variance, which makes it more complicated and interesting.

There are two principal types of SG, the Edwards-Anderson model [4] and the Sherrington-Kirkpatrick model [5]. In the latter all spins interact with all spins, and not just with nearest neighbors as in the Ising model.

A mathematical trick used to study SG systems is replica symmetry [4, 7], where a large number of identical systems are averaged over to calculate free energy and thermodynamic properties. In this case the spins acquire one more index to indicate the particular replica to which it belongs:

$$H = -\sum_{i,k,a} J_{ik}\, s^a_i\, s^a_k - \sum_{i,a} h\, s^a_i \tag{2}$$

Extra indices are also introduced when, unlike the Ising case, the spin is a multi-component vector, as in the two-dimensional (X-Y) model and the three-dimensional Heisenberg model.

## III. Social Clusters and SG

In the context of our social network many of the features of the usual physical SG are quite apt, but some need further generalization, because the social dynamics is more complex than that of an assembly of identical spins.

The utility function may represented by the negative of a spin-glass form Hamiltonian

$$H = - \sum_{i,k,a,b} J^{ab}_{ik} s^a_i s^b_k - \sum_{i,a} h^a_i s^a_i \qquad (3)$$

Here $s^a_i$ is the orientation of the *i*-th agent with respect to attribute *a*, $J^{ab}_{ik}$ couples such preferences of different attributes of different agents with assigned weights which may change, at least partially, randomly like a spin-glass, and/or may vary with learning and adaptation. The variable $h^a_i$ is an external field which may represent inherited or nature-induced preference of the agent *i* for characteristic *a*.

If there are *n* agents and *m* attributes we have a total of *m*X*m* matrices each of size *n*X*n*. This includes also self-interaction of different attributes of agents -$J^{ab}_{ii} s^a_i s^b_i$, which is absent in the physical case, because it would trivially add the fixed magnitude of each orientation vector in that case. In the socio-economic context this term is quite important as it adds the problem of the need for self-consistency of different attributes of any agent without involving other agents. The *J* coupling matrices are not in general diagonal and different attributes of different agents should be allowed to interact. For example, the preference of an agent for a particular commodity such as an expensive car may also affect the choice of a neighbor for another commodity which too is expensive.

The interaction with nature or an agent's genetic make-up is represented by the second term. Analogous to an external magnetic field we have a factor that tries to orient each agent, but the magnitude of the field now depends both on the agent and the attribute nature wants to influence, giving it two indices.

We have already noted that for different agents and different attributes, some links may have positive values (bandwagon effect - trying to align agents in the same direction where co-operative orientation is more satisfying [8]) and some may have negative values (where, as in a minority game, it is more satisfactory to be different). Hence the ferro-antiferro mixture creates a SG-type system. However, unlike an Ising SG with only +1 and -1 values of the spin, in the socio-economic context we have to enlarge the domain of choices considerably, say from –*s* to +*s*, as is the case for the measured component of a three-dimensional spin vector *s* in quantum mechanics. Later in this paper we shall also consider the continuum limit.

Unlike a usual SG model, it may be more appropriate in a socio-economic system to have nonzero average <*J*> couplings because the couplings may have fixed nonzero biases. We have commented on the impossibility of quantifying exactly all the factors that build up the utility function (or the concept of 'self' of an agent in a non-economic context) in terms of a manageably small number of attribute parameters. The interaction of undefined remaining parameters may be simulated by adding to the fixed bias matrix elements a random part, whereas in a usual SG the whole coupling is random. The presence of this element of randomness indicates the possibility of such a system having some usual features of SG.

The attribute index (superscript) is similar to the different components of a vector, but as *J* depends on it too, in nontrivial cases we would not get a normal scalar product between the spins as in a Heisenberg magnet. Though one can formulate an elliptical system with broken rotational for Heisenberg-type magnets, in this case as the sign of the *J* components can also change, we have effectively a noncompact attribute space which makes any comparison with Heisenberg magnets almost irrelevant. If we try to interpret these indices as replica indices, then too we have such grossly broken replica symmetry on account of the varying J's, that the analogy, though intriguing, cannot serve any useful purpose.

The total preference make-up of an agent or his 'self' should involve an extremely large number of attributes. This problem is already quite complex, more so than the physical models with small dimensionality, and hence it necessitates the selection of the most important attributes, i.e. those with the highest link weights $J^{ab}_{ik}$ to keep the dimensionality small.

The reduction of variables is also justified on grounds other than those of mathematical expediency. Even if we start with the assumption that a perfect definition of an agent at a certain time point may need infinitely many variables and preferences, the perception of "self" is limited with only a finite number of variables at a certain time point, as realizing a genetic variable and optimizing a need requires the expenditure of time and energy, and each agent is endowed with only a limited amount of such. As a result, at each time point, an agent can be approximated to a finite number of variables with high weight terms in the array.

Some of the weights are difficult to shift and are always high [9]. In an abstract sense, we can say that there is a variable called existence, which, when flipped, an agent ceases to exist, and some variables are attached to that with a high non-modifiable or difficult to update weight. For example, the utility eating is difficult to shift weight from. If an agent does not consider it important to eat, he or she will cease to live [10]    .

Again, some of these variables have a high weight because it can be thought that the agent is connected with the environment with an inflexible weight regarding those variables. The environment can be thought of as a heat bath the agent is in contact with. Hence, the variables of the environment can be taken as a thermodynamic average of a large ensemble, and are more fixed relative to the variables an individual agent possesses. Hence, it is safe to assume that agent-agent interactions can more easily flip variables than can agent-environment interactions.

**IV. Simulation of a Small World**

An exact analytic solution is possible for the simple Ising model and that too only for dimension 2. For a heterogeneous system with random component in the links like the one described above, simulation may be the only realistic method of studying the system. But even simulations become extremely time-consuming for large values of the number of agents or attributes. We have tried in this work to observe the dynamics of a world with three agents only, and each agent having three attributes only. The utility function is described by the negative of the explicit Hamiltonian

$$H = -\sum_{i,k,a,b} A^{ab}_{ik} s^a_i s^b_k - \sum_{i,a,b} B^{ab}_i s^a_i s^b_i - \sum_{i,a} h^a_i s^a_i \tag{4}$$

with nine *A* matrices coupling different attributes of different agents, three *B* matrices coupling different attributes of each agent, and one *h* matrix coupling each agent with the environment. For each agent, by turn, we update the value of the spin by evaluating

$$\frac{\partial H}{\partial s^a_i} = -\sum_{k,b} A^{ab}_{ik} s^b_k - \sum_b B^{ab}_i s^b_i - h^a_i \tag{5}$$

and updating $s^a_i$ by *-1* or *+1* according to whether this expression is greater or less than zero as each agent tries to maximize its utility, i.e. minimize *H*. The saturation value $|s| = s_{max}$ is taken to be *5*, so that even if Eq. 5 suggests a change to a higher $|s|$, we keep it fixed at the old value.

We experimented with different fixed components of the matrices *A, B* and *h*, and also with different relative weights of the fixed and random components. Among the elements in every coupling matrix we had fixed components of both positive and negative signs. The random components of the elements were given three different scales 1) 1X the magnitude as the fixed components, 2) 0.3 X the magnitude of the fixed components, and 3) 0.1X the magnitude of the fixed components to understand the relative importance of the fixed and random components in determining the qualitative dynamical behavior of the system. As the coupling matrices change slowly compared to the spin changes, the situation is similar to annealing.

## V. Results of Simulation

We have nine different evolutions for the nine $s^a_i$ variables and each of them was reproduced three times with the three different scales of randomness described above. However, an examination of all the developments shows a number of typical patterns, and we show the prototypes in Figs. 1-6.

In Fig. 1 we see an interesting feature in the time evolution of one agent's [#1] one attribute [#2]. It shows punctuated equilibria [11, 12], i.e. periods of static equilibrium punctuated by spurts of rapid changes. This phenomenon has been observed experimentally in geological evolution in fossil records. In social and economic contexts also periods of sudden changes interspersed with periods of stability are fairly common.

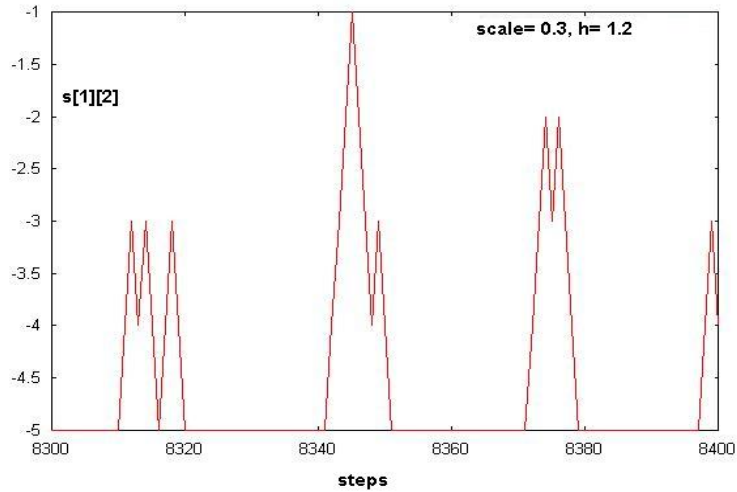

Fig. 1: Punctuated equilibrium for attribute 2 of agent 1 (short time development); randomness scale=0.3, external field = 1.2.

Fig.2 shows the behavior characteristic of most other $s^a_i$ variables, which is normal oscillation. Surprisingly, neither the frequency, nor the amplitude of the oscillations seem to show much fluctuation even when the randomness scale is large. It is well-known that the evolution of a coupled set of variables with opposite coupling signs yields an oscillation. In SG language, a frustrated system cannot find equilibrium and keeps oscillating.

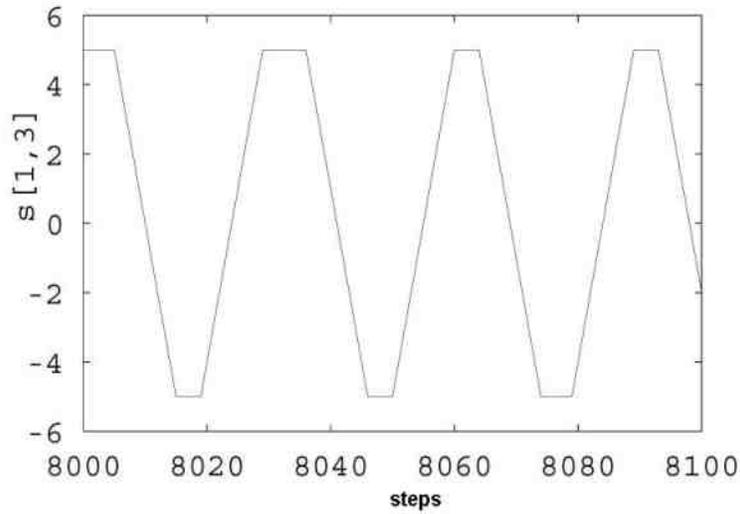

Fig. 2: Regular oscillations of attribute 3 of agent 1, for same scale and h.

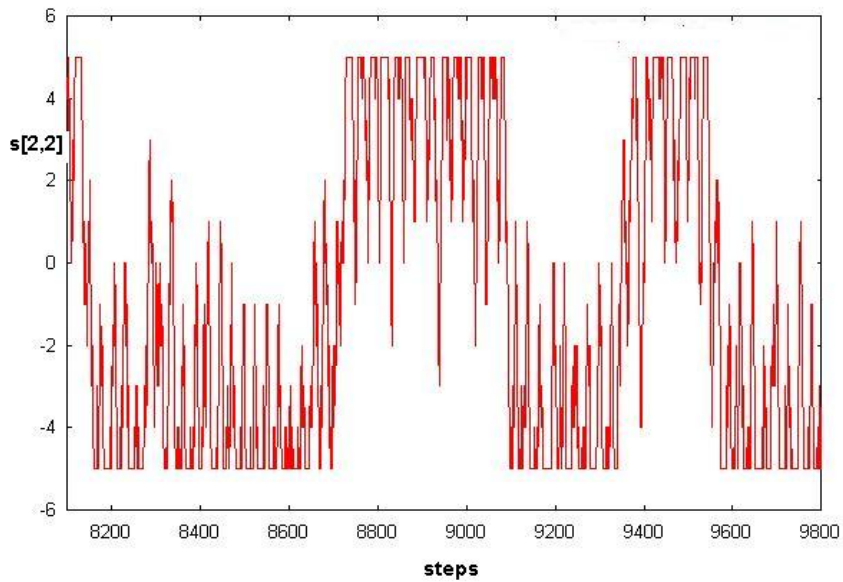

Fig. 3: Long term behavior change of s[2, 2], as base of short term oscillation flips suddenly

In Fig 3, we notice unexpected large time scale changes. It appears that the base of oscillations can reverse signs fairly rapidly so that there is a square wave oscillation with a much larger period than the period of short term oscillations. So, even in the simple example considered, we see the origin of a second time period from the dynamics of the system as a whole as distinct from the simple short time oscillations resulting from more direct conflicting interactions.

In Fig.4 we note a more gradual shift of base of short–term oscillations, which forms longer term oscillating trends. This is a familiar picture in socio-economic contexts, for example in stock-market indices.

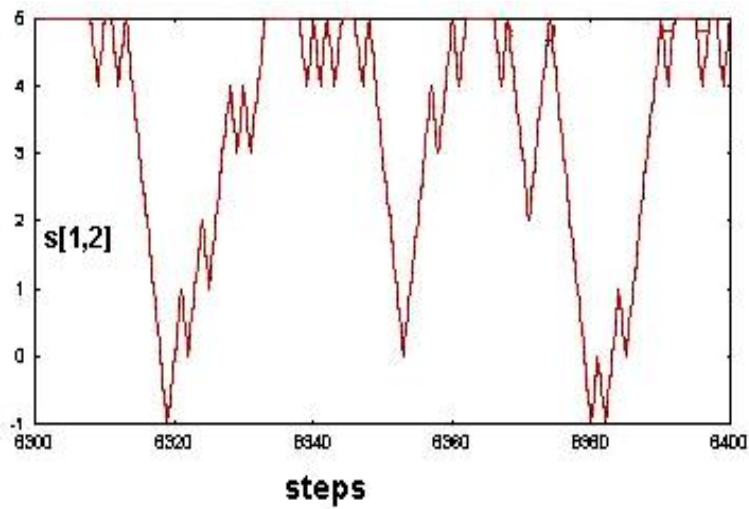

Fig. 4: Short period oscillations modulating longer period gradual reversals at random scale = 1.0, h= 0.

It is also possible to observe short-term of long-term near chaotic behavior for the same attribute of the same agent ( $s^2_1$ ) for changed values of the randomness scale and the external field as shown in Figs. 5 and 6.

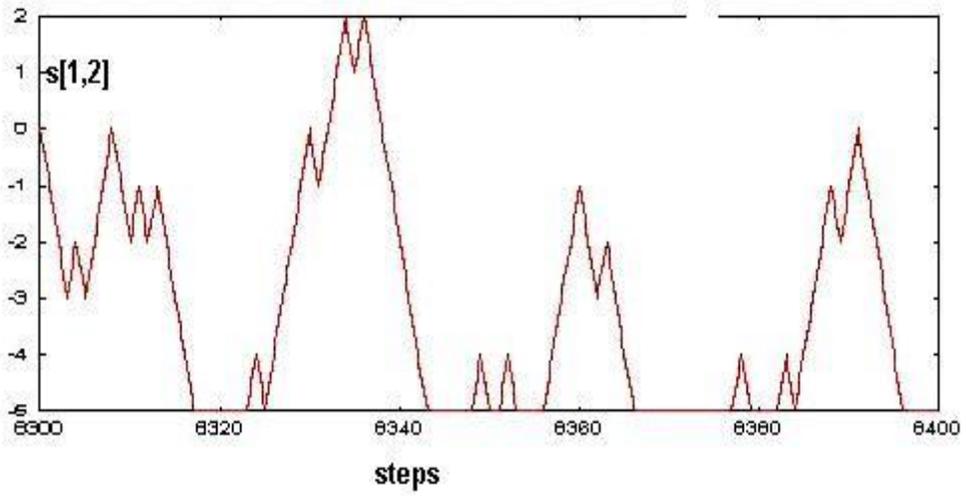

Fig. 5: Near chaotic behavior of s[1, 2] for randomness scale 0.3, h=0.

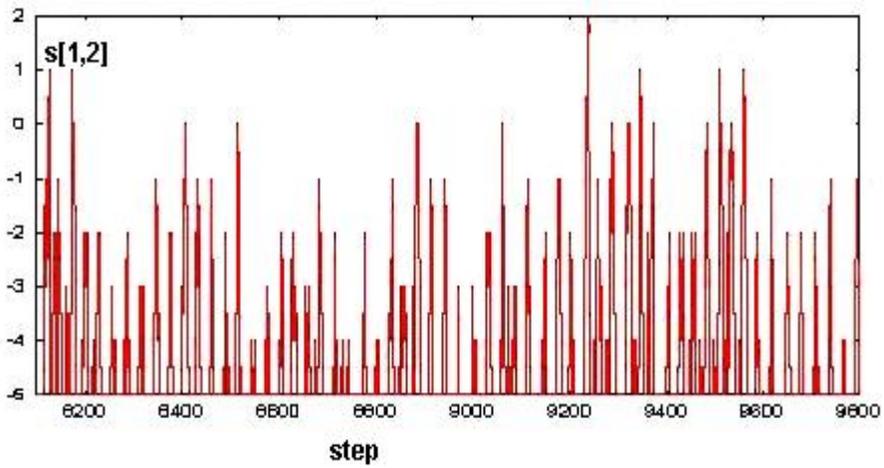

Fig. 6: Near chaotic long-term behavior for s[1, 2] for randomness scale = 0.3, h= 1.2

In Table I we summarize the different patterns of behavior for our choice of link matrices and for different random component scales and external filed values.

## TABLE I

*Summary of long and short term behavior for changes in randomness and external field*

| RANDOMNESS | EXTERNAL FIELD | SHORT TERM BEHAVIOR | LONG TERM BEHAVIOR |
|---|---|---|---|
| Small (0.1X fixed links) | 0 | Punctuated. Equil. | Nothing new |
| | 1.2 | Punctuated. Equil. | Nothing new |
| Medium (0.3X) | 0 | Irregular | Reversals |
| | 1.2 | Punctuated. Equil. | Irregular Amplitudes |
| Comparable (1X) | 0 | Irregular | Reversals |
| | 1.2 | Irregular | Reversals |

A simple method of observing the correlation between the same attribute of the different agents, which is important in socio-economic systems, may be the difference between the corresponding spin components. In Figs. 7 and 8 we see that in our model between the same two agents one attribute has a quasi-chaotic difference, and another is quite regular.

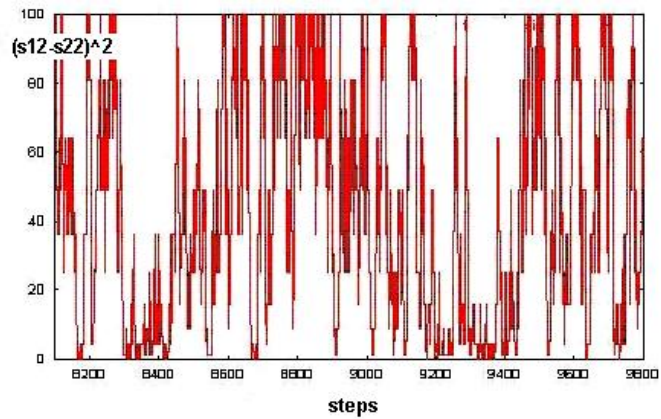

Fig. 7: $(s^2_1 - s^2_2)^2$ is fairly chaotic, showing little correlation between the agents for attribute *2*.

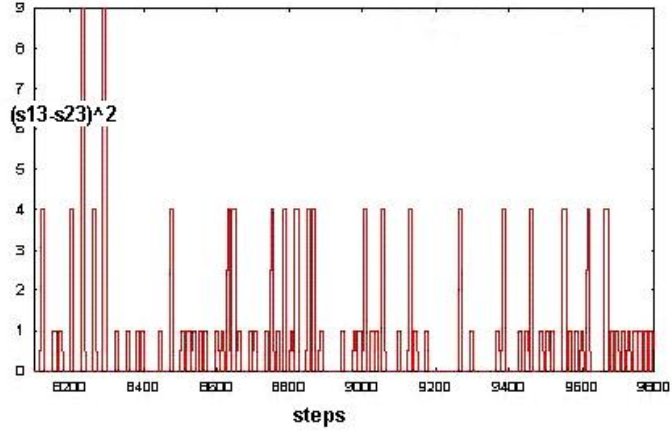

Fig. 8: $(s^3_1 - s^3_2)^2$ is small and regular, indicating good correlation between the same agents for attribute *3*.

## VI. Coupling as a Metric in State Space

If we use the combined notation $A= (a, i)$, $B= (b, j)$, then the interaction may be written as

$$H = - g_{AB} X^A X^B \tag{6}$$

where we have now made the "spin" (orientation) domain of the spin variable very large and continuous and have used the integrted vector symbol $X^A$ for it. This does not mean we are interested in actually allowing an agent to have infinite preference for any attribute, because the coupling $g_{AB}= J_{AB}$ can be made a function of $X^A$, and made to vanish asymptotically for high values of any component of $X^A$.

Now, $g_{AB}$ can be treated as a metric, and hence if we treat $X^A$ as the co-ordinate of the system in state space ( it contains the preference of every agent for every item/commodity in the system). As *H* in Eq. 6 is now similar to the distance function of differential gemometry we can write the equivalent "action" integral as the time integral of the Lagrangian [13]

$$L = \sqrt{-g_{AB} \frac{dX^A}{dt} \frac{dX^B}{dt}} \tag{7}$$

yielding the usual geodesic equation

$$\frac{d^2 X^A}{dt^2} + \Gamma^A_{BC} \frac{dX^A}{dt} \frac{dX^B}{dt} = 0 \tag{8}$$

where the Christoffel symbol is given by

$$\Gamma^A{}_{BC} = (1/2) g^{AD}( g_{BD,C} + g_{CD,B} - g_{BC,D}) \tag{9}$$

the subscripts after comma denoting partial differentiation with respect to the corresponding component of *X*.

Eq.8 relates the acceleration of change of state to a force coming from the curvature of the state space, which is contained in the functional relationship of the metric *g* (coupling *J*) on the state vector *X*. For a constant metric this is zero, giving inertial effects full control. In such a situation the preference variables $X^A$ may run away to infinity, which is not possible in a socio-economic system.

A simple compactification of the space in terms of projective space or a multidimensional sphere would not work, because it is not possible to identify points by going in opposing directions in a socio-economic system. However, as we have mentioned, $g_{AB}$ may contain damping factors for any large component of $X^A$. The situation reminds us of Fisher-Rao [14] metrics related to probability distribution functions (pdf). If the pdf is parametrized by the vector $a^A$, then a dispersion from the optimal value gives a distance between the new pdf and the old one as :

$$ds^2 = g_{AB} \, da^A \, da^B \tag{10}$$

where

$$g_{AB} = \frac{d^2 \log p}{da^A \, da^B} \tag{11}$$

The use of logarithm is justified by the functional relationship between probabilty and information or entropy. In a socio-economic context changes are meanignful almost always only in terms of fractions (percentage growth, literacy rate, employment, mortality etc.) so that the perceived value of a quantity is actually its logarithm, i.e. if we define

$$Y^A = \log X^A \tag{12}$$

Then a new definitionof utility in terms of these variables gives a more conservative metric

$$ds^2 = g_{AB} \, dY^A \, dY^B = g_{AB} \, (dX^A/X^A)(dX^B/X^B) \tag{13}$$

which tries to decouple attributes carrying high $X^A$ values, preventing run-away systems. The remaining $g_{AB}$ may contain further X dependence.

In a more elaborate model, a finite-temperature system may have a superposition of geometries with different metrics with corresponding probabilites.

## VII. Conclusions

We have seen that many of the concepts of the physical SG system are applicable to a socio-economic system, though generalizations going beyond strict physical constraints are also necessary. Our simulation of a small world with only three agents with three variable attributes which interact produced a surprisingly rich diversity of dynamical behavior and we have seen that both long and short term periodicties are noticeable as well as nearly chaotic patterns, which are the consequences of interaction between fixed and random components of the links between the agents and an external field representing nature.

Smallness of randomness in links gives greater regularity in the dynamic pattern, usually symmetric oscillations for some interesting attributes, and punctuated equilibrium for some attributes of some agents. For small randomness, long term behavior shows nothing new, but for large randomness long term behavior can show reversals of short term patterns at long intervals. The presence of an external field of same magnitude as links also imposes greater regularity, often with periodic long term patterns absent when, for the same randomness, there is no such field. Even when short term patterns look chaotic, well-defined long term patterns are often discernible. Correlation patterns are visible in some attributes, even when high randomness is present.

We have also seen that the state of a small system can be defined by a state vector and the coupling matrix then acts as a metric in the state space allowing us to investigate certain charactereistics of the system dynamics using the methods of differential geometry. A thermodynamic formulation may indicate the possibility of phase transitions, which is under investigation now.

The author would like to thank Professor P.W. Anderson and also Professor M.E. Fisher of the University of Maryland for encouragement.